\documentclass[aps,pra, twocolumn, amsmath, amssymb, showpacs, superscriptaddress]{revtex4}

\usepackage{graphicx}
\usepackage{bm}
\newcommand{\mr}{\mathrm}

\begin{document}
\title{Synthetic Frequency Protocol in the Ramsey Spectroscopy of Clock Transitions}

\author{V. I. Yudin}
\email{viyudin@mail.ru}
\affiliation{Novosibirsk State University, ul. Pirogova 2, Novosibirsk, 630090, Russia}
\affiliation{Institute of Laser Physics SB RAS, pr. Akademika Lavrent'eva 13/3, Novosibirsk, 630090, Russia}
\affiliation{Novosibirsk State Technical University, pr. Karla Marksa 20, Novosibirsk, 630073, Russia}
\author{A. V. Taichenachev}
\affiliation{Novosibirsk State University, ul. Pirogova 2, Novosibirsk, 630090, Russia}
\affiliation{Institute of Laser Physics SB RAS, pr. Akademika Lavrent'eva 13/3, Novosibirsk, 630090, Russia}
\author{M. Yu. Basalaev}
\affiliation{Novosibirsk State University, ul. Pirogova 2, Novosibirsk, 630090, Russia}
\affiliation{Institute of Laser Physics SB RAS, pr. Akademika Lavrent'eva 13/3, Novosibirsk, 630090, Russia}
\author{T. Zanon-Willette}
\affiliation{LERMA, Observatoire de Paris, PSL Research University, CNRS, Sorbonne Universit$\acute{e}$s, UPMC Univ. Paris 06, F-75005, Paris, France}

\date{\today}

\begin{abstract}
We develop an universal method to significantly suppress probe-induced shifts in any types of atomic clocks using the Ramsey spectroscopy. Our approach is based on adaptation of the synthetic frequency concept [V. I. Yudin, {\em et al}., Phys. Rev. Lett. {\bf 107}, 030801 (2011)] (previously developed for BBR shift suppression) to the Ramsey spectroscopy with the use of interrogations for different dark time intervals. Universality of the method consists in arbitrariness of the possible Ramsey schemes. However, most extremal results are obtained in combination with so-called hyper-Ramsey spectroscopy [V. I. Yudin, {\em et al}., Phys. Rev. A {\bf 82}, 011804(R) (2010)]. In the latter case, the probe-induced frequency shifts can be suppressed considerably below a fractional level of 10$^{-18}$ practically for any optical atomic clocks, where this shift previously was metrologically significant. The main advantage of our method in comparison with other radical hyper-Ramsey approaches [R. Hobson, {\em et al}., Phys. Rev. A {\bf 93}, 010501(R) (2016); T. Zanon-Willette, {\em et al}., Phys. Rev. A {\bf 93}, 042506 (2016)] consists in much greater efficiency and resistibility in the presence of decoherence.
\end{abstract}

\pacs{32.70.Jz, 06.30.Ft, 32.60.+i, 42.62.Fi}

\maketitle

At the present time, huge progress occurs for high-precision optical atomic clocks based on both neutral atoms in optical lattices \cite{katori2015,ye2015,oates2013,bize2013,ye2014,oates2014,sterr2014,ertmer2015} and trapped ions \cite{huntemann2016,rosenband2010,madej2012,gill2014}. Exceptional accuracy and stability at the
10$^{-17}$-10$^{-18}$ level are achieved. Potential possibilities to achieve the level of 10$^{-19}$ become clearer for nuclear clocks \cite{peik2003,campbell2012,yamaguchi2015,tkalya2015} and for highly charged ions \cite{derevianko2012,safronova2014prl,yudin2014}.
Great fundamental (e.g., in tests of fundamental physical theories such as QED, QCD, unification theories, cosmology, dark matter searches, etc.) and practical (navigation and information systems, gravity-geopotential surveying) importance of the current and long-range researches is  well-known and unquestionable. Current state, concomitant problems, and future prospects are well presented in the review \cite{ludlow2015rmp}.

On the way to these remarkable achievements, different barriers arise, which require the development of new unconventional approaches. As an example, for some of the promising clock systems, one of the key
problems is the frequency
shift of the clock transition due to the excitation pulses
themselves. For the case of magnetically induced spectroscopy \cite{yudin06,bar06} these
shifts (quadratic Zeeman and ac-Stark shifts) could ultimately limit the
achievable performance. Moreover, for ultra-narrow transitions (e.g., electric octupole \cite{hos09}
and two-photon transitions \cite{fis04,badr06}) the ac-Stark shift can
be so large in some cases to rule out high accuracy clock performance at all. A similar limitation
exists for clocks based on direct frequency comb spectroscopy \cite{fortier06,stowe08} due to ac-Stark shifts induced by large numbers of off-resonant laser modes.

Unconventional solution to this important problem was proposed in the paper \cite{yudin2010}, in which so-called hyper-Ramsey method has been developed. Soon this approach was successfully realized in \cite{hunt12}, where the huge suppression (by four orders of magnitude) of probe-induced shifts was experimentally demonstrated (see also \cite{huntemann2016}). However, a potential of this method was not going to be settled. In the experimental-theoretical paper \cite{NPL2015} a `stunning' result was recently shown: certain simple modification allows, in principle, {\em totally}(!) to exclude probe-induced shifts. Other hyper-Ramsey modification, having the same efficiency, was very soon proposed in the theoretical paper \cite{Zanon}. Because these phenomenal results can have far-reaching consequences for development of atomic clocks, it requires utterly thorough investigation of the schemes \cite{NPL2015,Zanon}. Besides, undoubted importance has a search of new variants to suppress probe-induced shifts with the near extremal efficiency.

In this paper, we develop an universal method to dramatically suppress probe-induced shifts and their fluctuations in any type of atomic clocks. Our approach is based on adaptation of so-called synthetic frequency concept \cite{yudin2011} to the Ramsey spectroscopy with the use of interrogations for different durations of free evaluation intervals. We show that this protocol in combination with the original hyper-Ramsey scheme \cite{yudin2010} makes most extremal results and is quite stable with respect to the decoherence. Moreover, our method leads to the much better and robust suppression of the shifts in comparison with protocols \cite{NPL2015,Zanon}, which are more sensitive to the decoherence and therefore do not show phenomenal results already.

\section{General theory}

The essence of our approach consists in the following. Previously in the paper \cite{yudin2011} the so-called synthetic frequency method, allowing to radically suppress thermal (BBR) shift in atomic clock, was proposed. However, an ideology of this method can be easy extended on cancelling of arbitrary systematic shift. Indeed, let us consider two clock frequencies $\omega^{(0)}_1$ and $\omega^{(0)}_2$ (different in the general case). Assume that due to a certain physical cause we have the stabilized frequencies $\omega^{}_1$ and $\omega^{}_2$, which are shifted relative to the unperturbed frequencies at the values $\Delta_{1}$ and $\Delta_{2}$:
\begin{equation}\label{initial}
\omega^{}_1=\omega^{(0)}_1+\Delta^{}_1;\quad \omega^{}_2=\omega^{(0)}_2+\Delta^{}_2.
\end{equation}
Also assume that the ratio $\varepsilon^{}_{12}$=$\Delta^{}_1/\Delta^{}_2$=$const$ does not fluctuate, while the shifts $\Delta_{1,2}$ can be varied during experiment (i.e., $\Delta_{1,2}$$\neq$$const$). In this case, we can construct the following superposition:
\begin{equation}\label{syn_gen}
\omega^{}_\text{syn}=\frac{\omega^{}_1-\varepsilon^{}_{12}\omega^{}_2}{1-\varepsilon^{}_{12}}= \frac{\omega^{(0)}_1-\varepsilon^{}_{12}\omega^{(0)}_2}{1-\varepsilon^{}_{12}},
\end{equation}
which is insensitive to the perturbations $\Delta_{1,2}$ and their fluctuations. This frequency we will call as `synthetic frequency'. A key advantage of this concept consists in the following: to construct the shift-free frequency $\omega^{}_\text{syn}$ we do not need to know the real values of shifts $\Delta_{1,2}$, because we need to know only their ratio $\varepsilon^{}_{12}$, which can be exactly calculated (or measured) for many cases.

Let us show how to incorporate the synthetic frequency protocol in the Ramsey spectroscopy for significant suppression of probe-induced shifts in atomic clocks. The general idea can be understandable from the reasonings related to the two-level system with unperturbed frequency $\omega^{}_{0}$. It is well-known that the standard Ramsey spectroscopy \cite{rams1950} uses two exciting pulses of resonance field with the frequency $\omega$, which are separated by the free evolution interval $T$ (dark time) [see in Fig.~\ref{fig:scheme}(a)]. In this case, the spectroscopic signal has a functional dependence on the detuning $\delta$=$\omega-\omega^{}_{0}$, which consists of set of narrow resonances with width of order of $\pi/T$ (so-called Ramsey fringes). The central fringe can be used as reference point for stabilisation of frequency $\omega$ in atomic clock.

Consider an influence of probe-induced shift $\Delta_\mr{sh}$, which arises only during the Ramsey pulses [see two-level scheme in Fig.~\ref{fig:scheme}], while this shift is absent during the dark time $T$. As a result, the stabilized frequency $\omega^{}_{T}$ also becomes differing from unperturbed frequency $\omega^{}_{0}$:
\begin{equation}\label{omega_shift}
\omega^{}_{T}=\omega^{}_{0}+\bar{\delta}^{}_{T}\,,
\end{equation}
where the index $T$ denotes the fixed time of the free evolution interval under frequency stabilization, and the resulting shift $\bar{\delta}^{}_{T}$$\neq$0 exists due to the $\Delta_\mr{sh}$$\neq$0. On the basis of general principles, it can be shown that the dependence $\bar{\delta}^{}_{T}$ on the value $T$ can be expressed as the following decreasing series in terms of powers of $1/T$:
\begin{equation}\label{shift_gen}
\bar{\delta}^{}_{T}=\frac{A_1}{T}+\frac{A_2}{T^2}+...+\frac{A_n}{T^n}+...\,,
\end{equation}
where the coefficients $A_n$ depend on the pulse parameters (durations, amplitudes, phases, and the value $\Delta_\mr{sh}$).

Because the time $T$ is precisely controlled in experiments, then we can set a goal to eliminate the main contribution $\propto A_1/T$ in Eq.~(\ref{shift_gen}) using the synthetic frequency protocol. To solve this task we will apply two different dark intervals $T_1$ and $T_2$ (but with the same Ramsey pulses), which will give us the corresponding stabilized frequencies $\omega^{}_{T_1}$ and $\omega^{}_{T_2}$. Using Eqs.~(\ref{syn_gen})-(\ref{shift_gen}) we easy find the synthetic frequency $\omega^{(1)}_\text{syn}$ and its residual shift $\bar{\delta}^{(1)}_\text{syn}$:
\begin{align}\label{syn_1}
&\omega^{(1)}_\text{syn}=\frac{\omega^{}_{T_1}-(T_2/T_1)\,\omega^{}_{T_2}}{1-(T_2/T_1)}\,, \nonumber\\
& \bar{\delta}^{(1)}_\text{syn}=\omega^{(1)}_\text{syn}-\omega^{}_{0}= \frac{\bar{\delta}^{}_{T_1}-(T_2/T_1)\,\bar{\delta}^{}_{T_2}}{1-(T_2/T_1)}\,,
\end{align}
where the expression for $\bar{\delta}^{(1)}_\text{syn}$ does not contain the term $\propto$$A_1$. For determinacy, below we will investigate in detail the particular case of $T_1$=$T$ and $T_2$=$T/2$:
\begin{align}\label{syn_1_1}
&\omega^{(1)}_\text{syn}=2\omega^{}_{T}-\omega^{}_{T/2}\,,\nonumber\\
&\bar{\delta}^{(1)}_\text{syn}=\omega^{(1)}_\text{syn}-\omega_{0} =2\bar{\delta}^{}_{T}-\bar{\delta}^{}_{T/2}\,.
\end{align}
As it will be shown below, the value $\bar{\delta}^{(1)}_\text{syn}$ can be less than $\bar{\delta}^{}_{T}$ [see Eq.~(\ref{shift_gen})] by several orders of magnitude.

Moreover, we can go further to define other synthetic frequency $\omega^{(2)}_\text{syn}$, for which both contributions $A_1/T$ and $A_2/T^2$ will be simultaneously canceled. Here we need to use three different time intervals ($T_1$, $T_2$, $T_3$) with the corresponding stabilized frequencies ($\omega^{}_{T_1}$, $\omega^{}_{T_2}$, $\omega^{}_{T_3}$). In particular, we will consider the case of $T_1$=$T$, $T_2$=$T/2$ and $T_3$=$T/3$, for which the required superposition takes the form:
\begin{align}\label{syn_2}
&\omega^{(2)}_\text{syn}=3\omega^{}_{T}-3\omega^{}_{T/2}+\omega^{}_{T/3}\,,\nonumber\\ &\bar{\delta}^{(2)}_\text{syn}=\omega^{(2)}_\text{syn}-\omega^{}_{0}=3\bar{\delta}^{}_{T}-3\bar{\delta}^{}_{T/2}+\bar{\delta}^{}_{T/3}\,.
\end{align}
As it will be shown below, the value $\bar{\delta}^{(2)}_\text{syn}$ can be less than $\bar{\delta}^{(1)}_\text{syn}$ [see Eq.~(\ref{syn_1_1})] by several orders of magnitude.

Using the same logic, the above procedure can be formally extended to the $n$-th order, when we will consider the synthetic frequency
$\omega^{(n)}_\text{syn}$ and corresponding residual shift $\bar{\delta}^{(n)}_\text{syn}$=$\omega^{(n)}_\text{syn}-\omega^{}_0$. In this case, the frequency $\omega^{(n)}_\text{syn}$ is a special superposition of ($n+1$) different stabilized frequencies ($\omega^{}_{T_1}$, $\omega^{}_{T_2}$,..., $\omega^{}_{T_n}$, $\omega^{}_{T_{n+1}}$) corresponding to the dark time intervals (${T_1}$, ${T_2}$,..., ${T_n}$, ${T_{n+1}}$). For $\bar{\delta}^{(n)}_\text{syn}$ the contributions ($A_1/T$, $A_2/T^2$,..., $A_n/T^n$) [see Eq.~(\ref{shift_gen})] are simultaneously canceled.

\section{Computational algorithm}

\begin{figure}[t]
\footnotesize
\includegraphics[width=8.5cm]{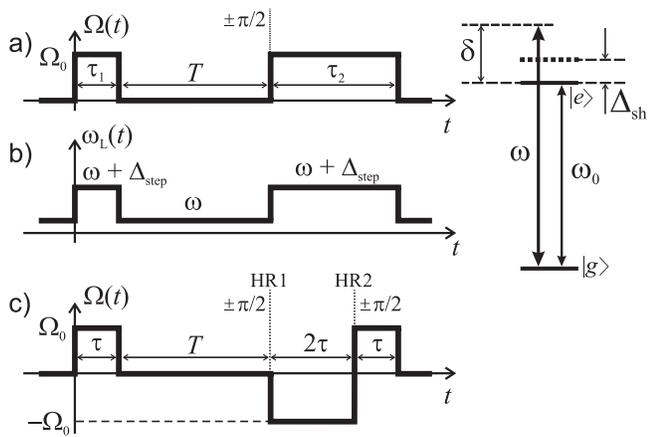}
\caption{Ramsey pulses with Rabi frequency $\Omega_0$ of different duration ($\tau_1$ and $\tau_2$; panel (a)). During the pulses, we step the laser frequency $\omega$ by $\Delta_{\rm step}$ (panel (b) and text). Hyper-Ramsey scheme with composite second pulse $3\tau$ (panel (c) and text). Also shown is a two-level atom with splitting $\omega_0$, detuning $\delta$ of the laser with frequency $\omega$ during dark time $T$, and excitation related shift $\Delta_\mr{sh}$ during pulses. }
\label{fig:scheme}
\end{figure}

Let us describe a computational algorithm, which allows us to calculate the signal in the Ramsey spectroscopy. The action of a single light pulse (with
frequency $\omega_\mr{p}$,  duration $\tau$, and Rabi frequency
$\Omega_0$) on two-level atoms with ground and excited states,
$|g\rangle$$=$${\small \left(\begin{array}{c} 0 \\ 1
\end{array}\right)}$ and $|e\rangle$$=$${\small \left(\begin{array}{c} 1 \\
0 \end{array}\right)}$ (separated by the unperturbed energy $\hbar \omega_0$), is described by the matrix:
\begin{eqnarray}\label{W}
&&\widehat{W}(\tau,\Omega_0,\delta_\mr{p})=\\
&&\begin{pmatrix}
    \cos\left(\frac{\Omega \tau}{2}\right)
                +\frac{i\delta_\mr{p}}{\Omega}\sin\left(\frac{\Omega \tau}{2}\right)&

    \frac{i\Omega_0}{\Omega}\sin\left(\frac{\Omega \tau}{2}\right) \\
    \frac{i\Omega_0}{\Omega}\sin\left(\frac{\Omega \tau}{2}\right) &
    \cos\left(\frac{\Omega \tau}{2}\right)
                -\frac{i\delta_\mr{p}}{\Omega}\sin\left(\frac{\Omega \tau}{2}\right)
\end{pmatrix},\nonumber
\end{eqnarray}
where $\Omega=\sqrt{\Omega_0^2+\delta_\mr{p}^2}$ is the generalized
Rabi frequency. The detuning during pulse
$\delta_\mr{p}=\omega_\mr{p}-\omega^{}_0-\Delta_\mr{sh}$ contains the
excitation related shift $\Delta_\mr{sh}$ (see
Fig.~\ref{fig:scheme}, level scheme) due to the influence of other
(far-off-resonant) transitions. Within the frequency interval corresponding to the narrow clock resonance the variation of $\Delta_\mr{sh}$ on $\omega_\mr{p}$ is negligible, i.e., $\Delta_\mr{sh}$ can be considered as a constant (for fixed $|\Omega_0|$).

During the dark period
between the pulses, excitation related shifts (which produce
the total actual shift $\Delta_\mr{sh}$) are absent (e.g., the ac-Stark shift
from the laser) or can be turned off (like the Zeeman shift). If
during the dark period $T$ the laser frequency is $\omega$, then
the free evolution is described by the matrix $\widehat{V}(T\delta)$ with detuning $\delta=\omega-\omega^{}_0$, where the matrix $\widehat{V}(x)$ is determined as:
\begin{equation}\label{V}
\widehat{V}(x)=
\left(\begin{array}{cc}
e^{ix /2}&0\\
0&e^{-ix /2}
\end{array}\right).
\end{equation}
In the general case, the laser frequency during the
pulse does not have to be the same as the frequency during the
dark time, i.e., $\omega_\mr{p} \neq\omega$ \cite{tai09}.  As we will see,
at times it can be useful to approximately offset the induced
shift, $\Delta_\mr{sh}$, by stepping the laser frequency only
during the pulses by a fixed $\Delta_\mr{step}$, i.e.,
$\omega_\mr{p} = \omega + \Delta_\mr{step}$ [see
Fig.~\ref{fig:scheme}(b)].  Thus, in the general case the detuning
during the pulses can be written as $\delta_\mr{p} = \delta -
\Delta$, where $\Delta = \Delta_\mr{sh} - \Delta_\mr{step}$ is the effective frequency shift during the pulse (instead of $\Delta_\mr{sh}$). This manipulation allows us to stabilise the frequency $\omega$ under controlled condition $|\Delta/\Omega_0|$$\ll$1 (independently of the value $\Delta_\mr{sh}$), which makes it possible to radically suppress the probe-induced shifts with the use of hyper-Ramsey spectroscopy \cite{yudin2010,hunt12} even for the large actual shifts $\Delta_\mr{sh}$: $|\Delta_\mr{sh}/\Omega_0|$$>$1. Indeed, if the actual level shift $\Delta_\mr{sh}$ is comparable to or
larger than $\Omega_0$, we can always apply a frequency step
$\Delta_{\rm step}$ (e.g., with an acousto-optic modulator) during
excitation to achieve the condition $|\Delta/\Omega_0| \ll 1$ for an
effective shift $\Delta$. $\Delta_{\rm step}$ can be evaluated
experimentally by variation of the dark period $T$. If $\Delta_{\rm
step} \neq \Delta_\mr{sh}$, the observed transition frequency will
be dependent on $T$ \cite{tai09}. With a control of the shift to 1\%
under typical conditions we can achieve $\left|\Delta/\Omega_0
\right| < 0.01$ to 0.1.

Formulas (\ref{W}) and  (\ref{V}) are sufficient for description of the signal in Ramsey spectroscopy. For example, if at $t=0$ atoms are in the lower level $|g\rangle$, then after
the action of two pulses of duration $\tau_1$ and $\tau_2$
separated by dark period $T$ (see Fig.~1a) the population
$n^{(e)}$ of atoms in the excited state $|e\rangle$ is determined by
\begin{equation}\label{eq:n_e}
n^{(e)}=\left|\langle e|
\widehat{W}(\tau_2,\Omega_0,\delta -\Delta)
\widehat{V}(T\delta)
\widehat{W}(\tau_1,\Omega_0,\delta-\Delta)
|g\rangle \right|^2.
\end{equation}
This formula describes Ramsey fringes as a function of
variable detuning $\delta$ (but with fixed $\Delta$). The presence
of the additional shift $\Delta$ in the course of the pulse action
leads to the shift of the central Ramsey fringe with respect to the unperturbed
frequency $\omega_0$.

\section{Synthetic frequency protocol for hyper-Ramsey spectroscopy}

Before the consideration of the synthetic frequency protocol, note some general points.
First of all, we assume that the position of the central fringe $\omega^{}_{T}$ is determined by stepping the phase of one of the pulses  by $\pm \pi/2$ in the way \cite{mor89} and equalizing these signals. This approach is of greater relevance for clocks, because it
directly generates an error signal with high sensitivity. In respect to the signal (\ref{eq:n_e}), this method is formulated as following. Let us introduce the phase steps $\phi$ after dark time $T$, which can expressed by the use of function $n^{(e)}_\text{R}(\phi)$ and the error signal $S^\text{(err)}_\text{R}$:
\begin{align}\label{ne_phi}
&n^{(e)}_\text{R}(\phi)= \nonumber\\
&\left|\langle e|
\widehat{W}(\tau_2,\Omega_0,\delta -\Delta)
\widehat{V}(\phi)\widehat{V}(T\delta)
\widehat{W}(\tau_1,\Omega_0,\delta-\Delta)
|g\rangle \right|^2,\nonumber\\
& S^\text{(err)}_\text{R}=n^{(e)}_\text{R}(\pi/2)-n^{(e)}_\text{R}(-\pi/2)\,.
\end{align}
Then the shift $\bar{\delta}^{}_{T}$ of stabilized frequency $\omega^{}_{T}$ is determined as solution of the equation $S^\text{(err)}_\text{R}=0$ relative to the unknown $\delta$.

At first, let us consider the standard Ramsey spectroscopy, in which both exciting pulses have the equal duration $\tau_1$=$\tau_2$=$\tau$.
In the case of $\Omega_0 \tau$=$\pi/2$ and $2\tau$$\ll$$T$, the dominating contributions have the following linear dependencies on the small value $|\Delta/\Omega_0|$$<$1:
\begin{align}\label{Syn_Rams}
&\bar{\delta}^{}_{T} \approx \frac{2}{T}\frac{\Delta}{\Omega_0};\;
\bar{\delta}^{(1)}_\text{syn} \approx \frac{8}{\pi T}\frac{2\tau}{T}\frac{\Delta}{\Omega_0};\; \bar{\delta}^{(2)}_\text{syn} \approx \frac{48}{\pi^2 T}\left(\frac{2\tau}{T}\right)^2\frac{\Delta}{\Omega_0}.
\end{align}
Here due to smallness of the ratio ($2\tau/T$)$\ll$1 (i.e., for short Ramsey pulses) we have the chain of inequalities: $|\bar{\delta}^{(2)}_\text{syn}|$$\ll$$|\bar{\delta}^{(1)}_\text{syn}|$$\ll$$|\bar{\delta}^{}_{T}|$. Thus, the synthetic frequency protocol can significantly suppress the shifts even for standard Ramsey spectroscopy.

However, most extremal results can be obtained by the use of hyper-Ramsey scheme \cite{yudin2010,zanon2015} imaged in Fig.~\ref{fig:scheme}(c). Here the main peculiarity is the composite pulse (with total duration $3\tau$), which consists of sub-pulse $2\tau$ with inverted phase ($-\Omega_0$) and sub-pulse $\tau$ with initial phase ($\Omega_0$). If for the error signal we apply additional phase $\pm \pi/2$-steps directly after dark time (as it was in \cite{yudin2010,hunt12}), then this method we will denote as HR1 [see in Fig.~\ref{fig:scheme}(c)]. In this case, we define the function $n_\text{HR1}^{(e)}(\phi)$ and the error signal $S_\text{HR1}^\text{(err)}$:
\begin{align}\label{n_HRA}
&n_\text{HR1}^{(e)}(\phi)=| \langle e|\widehat{W}(\tau,\Omega_0,\delta-\Delta) \nonumber\\
&\widehat{W}(2\tau,-\Omega_0,\delta-\Delta)
\widehat{V}(\phi)\widehat{V}(T\delta)
\widehat{W}(\tau,\Omega_0,\delta-\Delta)
|g\rangle |^2\,, \nonumber \\
& S_\text{HR1}^\text{(err)}=n_\text{HR1}^{(e)}(\pi/2)-n_\text{HR1}^{(e)}(-\pi/2)\,.
\end{align}
Then the position of the stabilized frequency $\omega^{}_{T}$ (i.e., shift $\bar{\delta}^{}_{T}$) is determined by the solution of equation  $S_\text{HR1}^\text{(err)}=0$ relative to the unknown $\delta$. As it was first shown in \cite{yudin2010}, for HR1 the dominating contribution in the shift $\bar{\delta}^{}_{T}$ has {\em cubic} dependence on the small value $|\Delta/\Omega_0|$$\ll$1.

\begin{figure}[t]
\footnotesize
\includegraphics[width=8.5cm]{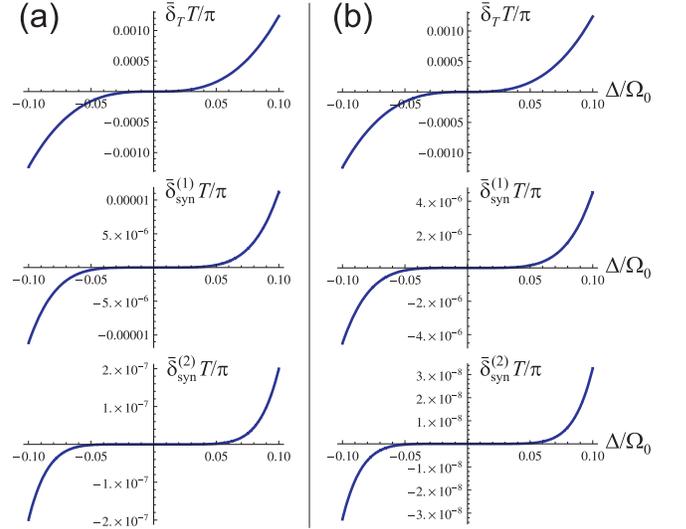}
\caption{The dependencies of the shifts $\bar{\delta}^{}_{T}$, $\bar{\delta}^{(1)}_\text{syn}$, and $\bar{\delta}^{(2)}_\text{syn}$ for HR1. Calculations are done for $\Omega_0 \tau$=$\pi/2$ and for different values $4\tau/T$: (a) $4\tau/T$=0.25; (b) $4\tau/T$=0.1. }
\label{HRA_1}
\end{figure}

In the Fig.~\ref{HRA_1} we demonstrate the calculations for HR1 without synthetic frequency [see $\bar{\delta}^{}_{T}$ in Fig.~\ref{HRA_1}], as well as with the use of synthetic frequency protocol [see $\bar{\delta}^{(1)}_\text{syn}$ and $\bar{\delta}^{(2)}_\text{syn}$ in Fig.~\ref{HRA_1}]. A huge advantage of the synthetic frequency protocol is obvious. Under $|\Delta/\Omega_0|^2$$\ll$1 our calculations show the following general character of the dominating dependencies on $\Delta/\Omega_0$:
\begin{align}\label{Syn_HRA_gen}
& \bar{\delta}^{}_{T} \propto \left( \frac{\Delta}{\Omega_0}\right)^3;\\
& \bar{\delta}^{(1)}_\text{syn} \propto \left( \frac{\Delta}{\Omega_0}\right)^5 ;\; \bar{\delta}^{(2)}_\text{syn} \propto  \left(\frac{\Delta}{\Omega_0}\right)^7;...;\, \bar{\delta}^{(n)}_\text{syn} \propto \left( \frac{\Delta}{\Omega_0}\right)^{2n+3}.\nonumber
\end{align}
Thus, for synthetic frequencies a higher-order (more than cubic) nonlinearities appear. Moreover, this character is not changed under variations of $\Omega_0$, $\tau$, and $T$, i.e., we absolutely do not need the rigorous condition $\Omega_0\tau$=$\pi/2$. This circumstance is a key point to successfully realize our method in atomic clocks, because in real experiments the value of $\Omega_0$ can be controlled only at the level of 1-10$\%$.

For instance, we present formulas under $\Omega_0 \tau$=$\pi/2$ and ($4\tau/T$)$<$1 [see also Appendix]:
\begin{align}\label{Syn_HRA}
& \bar{\delta}^{}_{T} \approx \frac{4}{T}\left( \frac{\Delta}{\Omega_0}\right)^3 ,\\
&\bar{\delta}^{(1)}_\text{syn} \approx \frac{48}{\pi T}\frac{4\tau}{T}\left( \frac{\Delta}{\Omega_0}\right)^5 ;\quad \bar{\delta}^{(2)}_\text{syn} \approx \frac{865}{\pi^2T}\left(\frac{4\tau}{T}\right)^2 \left(\frac{\Delta}{\Omega_0}\right)^7.\nonumber
\end{align}
These formulas demonstrate that the chain of inequalities, $|\bar{\delta}^{(2)}_\text{syn}|$$\ll$$|\bar{\delta}^{(1)}_\text{syn}|$$\ll$$|\bar{\delta}^{}_{T}|$, can be realized due to the controlled smallness $|\Delta/\Omega_0|^2$$\ll$1, first of all. Besides, the condition ($4\tau/T$)$\ll$1 (i.e., the use of short Ramsey pulses) leads to an additional suppression of the shifts. Thus, relatively small initial shift $\bar{\delta}^{}_{T}$ and its fluctuations can be dramatically suppressed (by several orders of magnitude) to the metrologically negligible values at all.

In the resent paper \cite{NPL2015}, authors have proposed the use of $\pm \pi/2$ phase steps after inverted ($-\Omega_0$) sub-pulse $2\tau$ [see in Fig.~\ref{fig:scheme}(c)] to form an error signal for hyper-Ramsey approach (we will denote this method as HR2). For its theoretical description we introduce the following function $n_\text{HR2}^{(e)}(\phi)$ and the error signal $S_\text{HR2}^\text{(err)}$:
\begin{align}\label{n_HRB}
&n_\text{HR2}^{(e)}(\phi)=| \langle e|\widehat{W}(\tau,\Omega_0,\delta-\Delta)\widehat{V}(\phi) \nonumber\\
&\widehat{W}(2\tau,-\Omega_0,\delta-\Delta)
\widehat{V}(T\delta)
\widehat{W}(\tau,\Omega_0,\delta-\Delta)
|g\rangle |^2\,,\nonumber\\
& S_\text{HR2}^\text{(err)}=n_\text{HR2}^{(e)}(\pi/2)-n_\text{HR2}^{(e)}(-\pi/2)\,.
\end{align}
Then the shift $\bar{\delta}^{}_{T}$ of stabilized frequency $\omega^{}_{T}$ is determined by the solution of equation $S_\text{HR2}^\text{(err)}=0$ relative to the unknown $\delta$. In this case, our theoretical estimations give us
practically the same expressions as Eqs.~(\ref{Syn_HRA}), but with opposite sign for $\bar{\delta}^{}_{T}$ and $\bar{\delta}^{(2)}_\text{syn}$:
\begin{align}\label{Syn_HRB}
&\bar{\delta}^{}_{T} \approx -\frac{4}{T}\left( \frac{\Delta}{\Omega_0}\right)^3,\\
&\bar{\delta}^{(1)}_\text{syn} \approx \frac{48}{\pi T}\frac{4\tau}{T}\left( \frac{\Delta}{\Omega_0}\right)^5 ,\quad \bar{\delta}^{(2)}_\text{syn} \approx -\frac{865}{\pi^2T}\left(\frac{4\tau}{T}\right)^2 \left(\frac{\Delta}{\Omega_0}\right)^7,\nonumber
\end{align}
under $\Omega_0\tau$=$\pi/2$ and $|\Delta/\Omega_0|$$\ll$1.

\section{Comparison between different hyper-Ramsey approaches in the presence of decoherence}

However, besides HR2 the paper \cite{NPL2015} also describes the theory and successful experimental demonstration of modified hyper-Ramsey (MHR) method, which is based on combination of both protocols HR1 and HR2. In the case of MHR, the error signal is determined as following
\begin{align}\label{MHR}
S_\text{MHR}^\text{(err)}=n_\text{HR1}^{(e)}(\phi)-n_\text{HR2}^{(e)}(\phi)\,.
\end{align}
Then the shift $\bar{\delta}^{}_{T}$ of stabilized frequency $\omega^{}_{T}$ is determined by the equation $S_\text{MHR}^\text{(err)}=0$
relative to the unknown $\delta$. This equation leads to an exceptional result: $\bar{\delta}^{}_{T}=0$ for {\em arbitrary} $\phi$, $\Delta$, $\Omega_0$, $\tau$, and $T$. An analogous result takes play also for alternative scheme [so-called generalized hyper-Ramsey (GHR)], presented in the theoretical paper \cite{Zanon}. Mathematical description of GHR [including the error signal $S_\text{GHR}^\text{(err)}$] can be expressed by the formulas:
\begin{align}\label{GHR}
&n_\text{GHR}^{(e)}(\phi)=| \langle e|\widehat{W}(\tau,\Omega_0,\delta-\Delta)\widehat{V}(-\phi) \nonumber\\
&\widehat{W}(2\tau,\Omega_0,\delta-\Delta)
\widehat{V}(\phi)\widehat{V}(T\delta)
\widehat{W}(\tau,\Omega_0,\delta-\Delta)
|g\rangle |^2;\nonumber\\
&S_\text{GHR}^\text{(err)}=n_\text{GHR}^{(e)}(\phi)-n_\text{GHR}^{(e)}(-\phi)\,,
\end{align}
where the shift $\bar{\delta}^{}_{T}$ is determined by the solution of equation $S_\text{GHR}^\text{(err)}=0$ relative to the unknown $\delta$. As it was shown in \cite{Zanon}, there is the same result: $\bar{\delta}^{}_{T}=0$ for arbitrary $\phi$, $\Delta$, $\Omega_0$, $\tau$, and $T$.

At first glance, both MHR and GHR aproaches \cite{NPL2015,Zanon} are absolutely ideal for the frequency stabilization, because they allow us totally to eliminate probe-induced shifts, i.e., the above concept of synthetic frequency protocol becomes not so important. However, as it will be shown below, MHR and GHR methods are unstable relative to the decoherence, which leads to an appearance of the shift $\bar{\delta}^{}_{T}$$\neq$0. Moreover, this residual shift is very sensitive to variations of Rabi frequency $\Omega_0$. At the same time, our approach is much more stable relative to the decoherence and it can be significantly better and robustly in real experiments than both MHR and GHR.

\begin{figure}[t]
\footnotesize
\includegraphics[width=8.5cm]{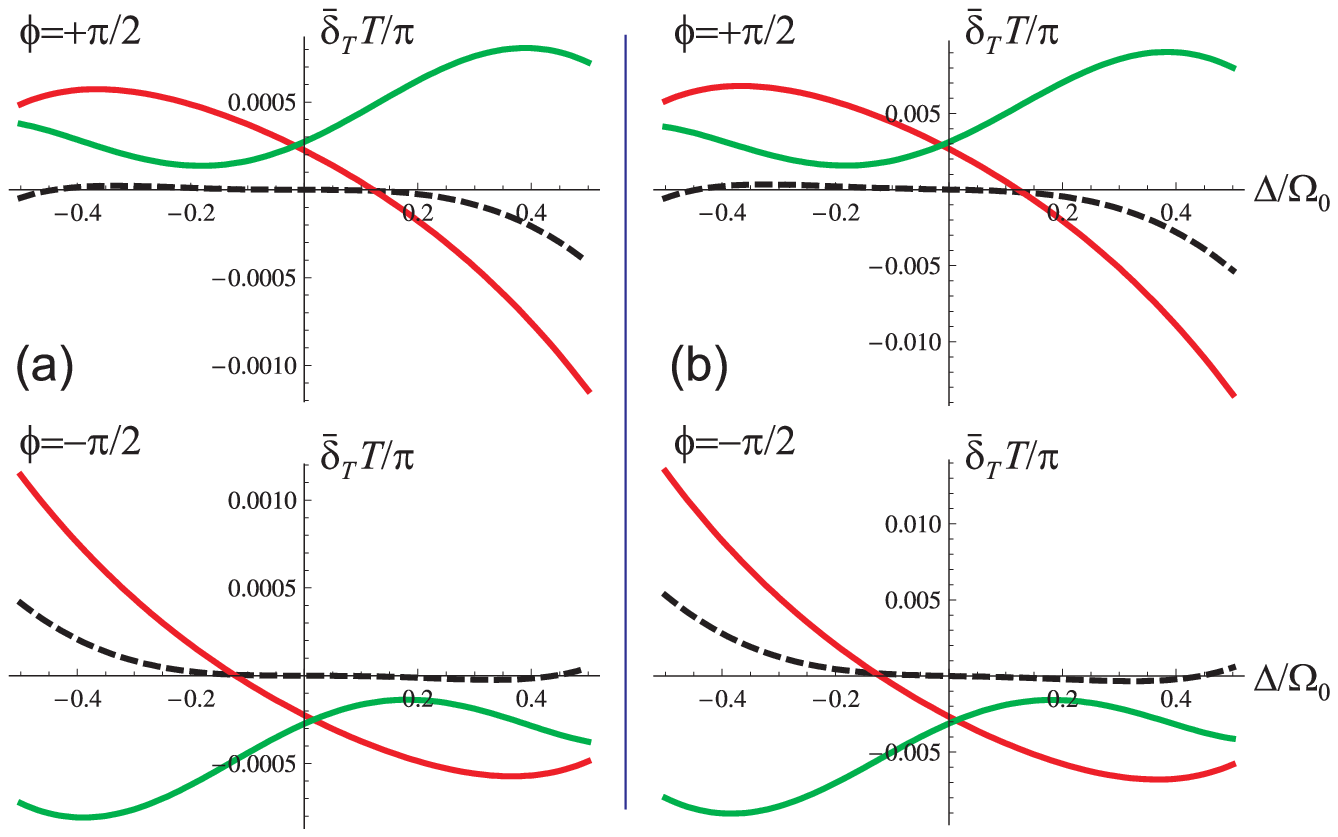}
\caption{The dependencies of shifts $\bar{\delta}^{}_{T}$ for MHR [see Eq.~(\ref{MHR})] under the decoherence ($\Gamma$$\neq$0). Calculations are done for $4\tau/T$=0.25, for different Rabi frequencies: $\Omega_0 \tau$=$\pi/2$ (black dashed lines); $\Omega_0 \tau$=0.9$\pi/2$ (red lines); $\Omega_0 \tau$=1.1$\pi/2$ (green lines), and for different $\Gamma$: (a) $\Gamma$=0.01$\pi/T$; (b) $\Gamma$=0.1$\pi/T$. Upper pictures are obtained for $\phi$=$+\pi/2$, and lower pictures are obtained for $\phi$=$-\pi/2$ [see Eq.~(\ref{MHR})]. One can see that upper and lower graphs correspond each other by the inversion relative to the central point (0,0).}
\label{MHR_comb}
\end{figure}

\begin{figure}[t]
\footnotesize
\includegraphics[width=8.5cm]{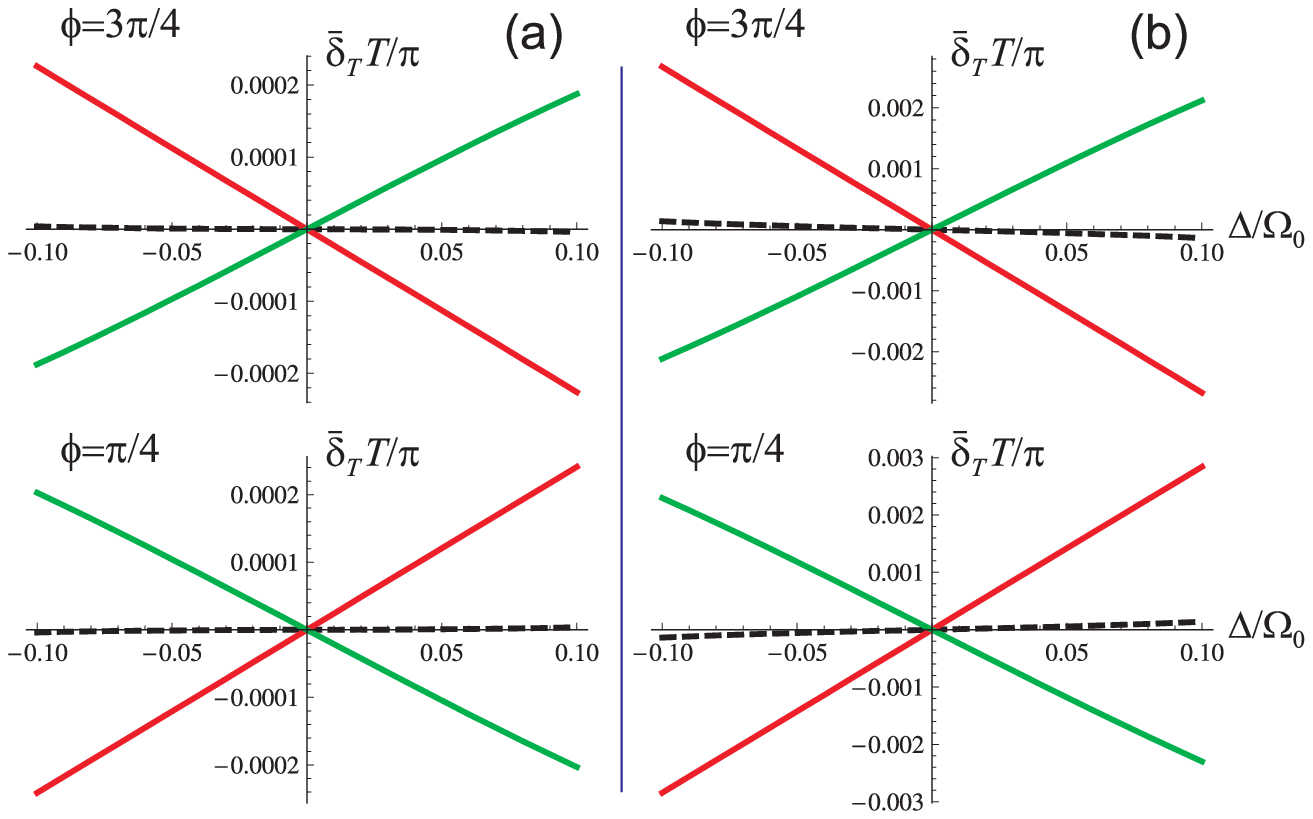}
\caption{The dependencies of shifts $\bar{\delta}^{}_{T}$ for GHR [see Eq.~(\ref{GHR})] under the decoherence ($\Gamma$$\neq$0). Calculations are done for $4\tau/T$=0.25, for different Rabi frequencies: $\Omega_0 \tau$=$\pi/2$ (black dashed lines); $\Omega_0 \tau$=0.9$\pi/2$ (red lines); $\Omega_0 \tau$=1.1$\pi/2$ (green lines), and for different $\Gamma$: (a) $\Gamma$=0.01($\pi/T$); (b) $\Gamma$=0.1($\pi/T$). Upper pictures are obtained for $\phi$=$3\pi/4$, and lower pictures are obtained for $\phi$=$\pi/4$ [see Eq.~(\ref{GHR})].}
\label{GHR_comb}
\end{figure}

To describe the Ramsey spectroscopy in the presence of decoherence, we will use the formalism of density matrix $\hat{\rho}$, which has the form
\begin{equation}\label{rho}
\hat{\rho}(t)=\sum_{j,k=g,e}|j\rangle \rho_{jk}^{}(t)\langle k|;\;\; \rho^{}_{gg}\equiv n^{(g)};\;\; \rho^{}_{ee}\equiv n^{(e)},
\end{equation}
in the basis of states $|g\rangle$ and $|e\rangle$. In our case, the density matrix components $\rho_{jk}^{}(t)$ satisfy the following differential equations:
\begin{align}\label{2_level}
&[\partial_t+\Gamma-i\tilde{\delta}(t)]\rho^{}_{eg}=i\Omega(t)[n^{(g)}-n^{(e)}]/2\,;\quad  \rho^{}_{ge}=\rho^{\ast}_{eg};\nonumber \\
&\partial_t n^{(e)}=i[\Omega(t)\rho_{ge}-\rho_{eg}\Omega^{\ast}(t)]/2\,;\;\; n^{(g)}+n^{(e)}=1.
\end{align}
Here the time dependencies $\Omega(t)$ and $\tilde{\delta}(t)$ are determined by the following: $\Omega(t)$=$\,\Omega_0$ (or $-\Omega_0$) and $\tilde{\delta}(t)$=$\,\delta-\Delta$ during the action of Ramsey pulses, but $\Omega(t)$=$\,0$ and $\tilde{\delta}(t)$=$\,\delta$ during the dark time $T$. The main difference of above equations from the Schr$\ddot{\text{o}}$dinger equation consists in the presence of the relaxation constant $\Gamma\,$$>$$\,0$ (for off-diagonal matrix elements $\rho^{}_{eg}$ and $\rho^{}_{ge}$), which describes the decoherence. In particular, such simple model allows us to estimate the influence of nonzero spectral width of the probe field. To achieve this goal, we can assume the order-of-magnitude agreement between the value $\Gamma$ and spectral width of the probe field. Similar estimations are very important, because even best modern lasers, used in atomic clocks, have the spectral width at the level of 0.1~Hz. Moreover, there are other possible causes of decoherence, which are connected with an action of environment, an influence of regimes of traps (or lattices), etc.

\begin{figure}[t]
\footnotesize
\includegraphics[width=8.5cm]{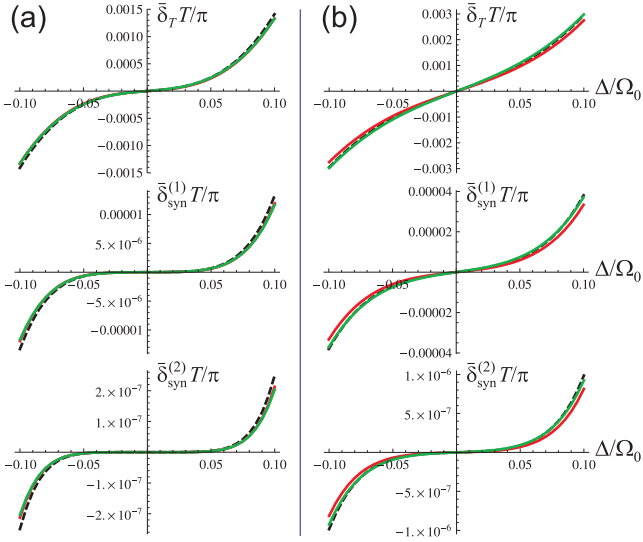}
\caption{The dependencies of shifts $\bar{\delta}^{}_{T}$, $\bar{\delta}^{(1)}_\text{syn}$, and $\bar{\delta}^{(2)}_\text{syn}$ for HR1 under the decoherence ($\Gamma$$\neq$0). Calculations are done for $4\tau/T$=0.25, for different Rabi frequencies: $\Omega_0 \tau$=$\pi/2$ (black dashed lines); $\Omega_0 \tau$=0.9$\pi/2$ (red lines); $\Omega_0 \tau$=1.1$\pi/2$ (green lines), and for different $\Gamma$: (a) $\Gamma$=0.01$\pi/T$; (b) $\Gamma$=0.1$\pi/T$. }
\label{HRA_comb}
\end{figure}

Fig.~\ref{MHR_comb} shows that in the presence of decoherence the MHR leads to the residual shift, which significantly depends on variations of Rabi frequency $\Omega_0$. Moreover, under condition $\Omega_0 \tau$$\neq$$\pi/2$ the method MHR makes `parasitic' shift $\bar{\delta}^{}_{T}$$\neq$0 even if $\Delta$=0. The next Fig.~\ref{GHR_comb} for GHR also demonstrates residual shifts and their strong sensitivity to the variation of value $\Omega_0$ under the decoherence. However, method GHR does not produce `parasitic' shift for $\Delta$=0. In contrast to both MHR and GHR, the synthetic frequency protocol, combined with original hyper-Ramsey scheme HR1, shows very good stability relative to both the decoherence and variations $\Omega_0$ [see $\bar{\delta}^{(1)}_\text{syn}$ and $\bar{\delta}^{(2)}_\text{syn}$ in Figs.~\ref{HRA_comb}(a),(b)]. While the `simple' hyper-Ramsey can be worse than MHR and GHR [compare $\bar{\delta}^{}_{T}$ in Fig.~\ref{HRA_comb} with $\bar{\delta}^{}_{T}$ in Figs.~\ref{MHR_comb},\ref{GHR_comb}].

\begin{figure}[t]
\footnotesize
\includegraphics[width=8.5cm]{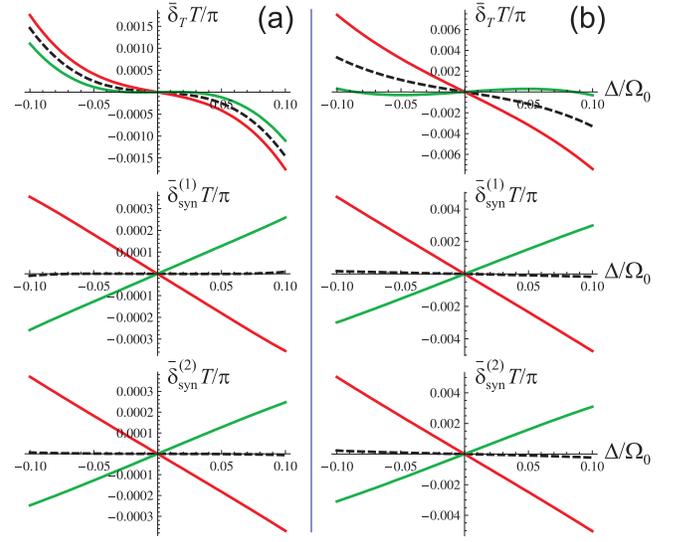}
\caption{The dependencies of shifts $\bar{\delta}^{}_{T}$, $\bar{\delta}^{(1)}_\text{syn}$, and $\bar{\delta}^{(2)}_\text{syn}$ for HR2 under the decoherence ($\Gamma$$\neq$0). Calculations are done for $4\tau/T$=0.25, for different Rabi frequencies: $\Omega_0 \tau$=$\pi/2$ (black dashed lines); $\Omega_0 \tau$=0.9$\pi/2$ (red lines); $\Omega_0 \tau$=1.1$\pi/2$ (green lines), and for different $\Gamma$: (a) $\Gamma$=0.01$\pi/T$; (b) $\Gamma$=0.1$\pi/T$. }
\label{HRB_comb}
\end{figure}

Thus, in the presence of decoherence, the synthetic frequency protocol in combination with HR1 \cite{yudin2010} can be much better (by one-three orders of magnitude) than both MHR \cite{NPL2015} and GHR \cite{Zanon}. Moreover, our calculations have shown an uselessness of synthetic frequency protocol for both MHR and GHR. It can be explained by the difference of the general dependence $\bar{\delta}^{}_{T}$ on parameter $T$ in comparison with Eq.~(\ref{shift_gen}): for MHR and GHR this dependence contains also constant contribution $A_0\propto\Gamma$. Besides, our calculations show [see Fig.~\ref{HRB_comb}] that the synthetic frequency protocol for HR2 is much worse (under the decoherence) than for HR1. Probably it can be explained in the following way: $\pm \pi/2$ phase steps for HR1 are directly conjugated with the dark interval $T$, while these $\pm \pi/2$ steps for HR2 are isolated from  the interval $T$ by the sub-pulse $2\tau$ (with inverted phase $-\Omega_0$) [see HR1 and HR2 in the Fig.~\ref{fig:scheme}(c)].

\subsection{Additional synthetic frequency protocol for GHR}
\begin{figure}[t]
\footnotesize
\includegraphics[width=8.5cm]{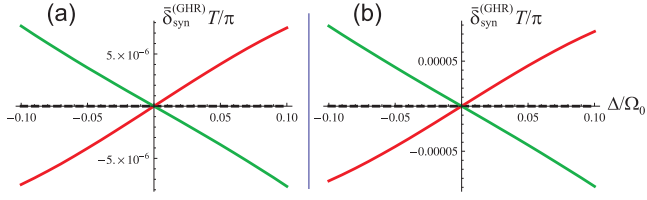}
\caption{The dependencies of shifts $\bar{\delta}^\text{(GHR)}_\text{syn}$ for synthetic frequency (\ref{GHR_syn_eq}) under the decoherence ($\Gamma$$\neq$0). Calculations are done for $4\tau/T$=0.25, for different Rabi frequencies: $\Omega_0 \tau$=$\pi/2$ (black dashed lines); $\Omega_0 \tau$=0.9$\pi/2$ (red lines); $\Omega_0 \tau$=1.1$\pi/2$ (green lines), and for different $\Gamma$: (a) $\Gamma$=0.01$\pi/T$; (b) $\Gamma$=0.1$\pi/T$. }
\label{GHR_syn}
\end{figure}

If we will more attentively look at Fig.~\ref{GHR_comb} and will compare upper and lower graphs, then we can see that these dependencies are practically transformed each to other under reversal of sign. It allows us for GHR to use other type of synthetic frequency, which is formed as half-sum of two frequencies:
\begin{equation}\label{GHR_syn_eq}
\omega^\text{(GHR)}_\text{syn}=\frac{1}{2}\left(\omega^{}_T|^{}_{\phi=3\pi/4}+\omega^{}_T|^{}_{\phi=\pi/4}\right)\,,
\end{equation}
for $\phi$=$3\pi/4$ and $\phi$=$\pi/4$ [see Eq.~(\ref{GHR})]. In this case, as seen from Fig.~\ref{GHR_syn}, the residual shift $\bar{\delta}^\text{(GHR)}_\text{syn}=\omega^\text{(GHR)}_\text{syn}-\omega^{}_0$ becomes much less than initial shifts $\bar{\delta}^{}_T$ in Fig.~\ref{GHR_comb}.

\section{Conclusion}

In our figures we use dimensionless quantities, because it allows us to use these calculations for different experimental conditions. In particular, the parameter $\bar{\delta}T/\pi$ corresponds to the ratio of the shifts ($\bar{\delta}^{}_{T}$, $\bar{\delta}^{(1)}_\text{syn}$, $\bar{\delta}^{(2)}_\text{syn}$) to the typical width (FWHM) of Ramsey resonances $\pi/T$ (in the s$^{-1}$ units). For example, if this width is equal to 10~Hz, then to suppress shifts to the level $<$0.1~mHz (i.e., to the fractional level less than 10$^{-18}$ for an optical range) we need fulfillment of ($\bar{\delta}T/\pi$)$<$10$^{-5}$. The same is related to the decoherence constant $\Gamma$: this value is also given on a scale of $\pi/T$ (see $\Gamma$=0.01$\pi/T$ or $\Gamma$=0.1$\pi/T$ in the captions). For example, the value $\Gamma$=0.01$\pi/T$ corresponds to the $\Gamma/2\pi$=0.1~Hz for 10~Hz width of Ramsey resonances.

In general, our calculations show that for $\Gamma/2\pi$$>$0.01-0.001~Hz the synthetic frequency protocol in combination with HR1 \cite{yudin2010} produces much more robust suppression of the probe-induced shifts to the fractional level of 10$^{-18}$-10$^{-19}$ in comparison with methods MHR \cite{NPL2015} and GHR \cite{Zanon}. Moreover, our approach allows us to achieve this level even for $(\Gamma/2\pi)$$>$$1$Hz. Apart from the combination with Ramsey and hyper-Ramsey spectroscopy for two-level systems, we have applied the synthetic frequency protocol to the Ramsey spectroscopy of the coherent population trapping (CPT) resonances (e.g., see \cite{zanon2005,chen2010,blanshan2015}). Note that CPT clock is one of perspective variants of compact rf clocks with relatively high metrological characteristics. Our calculations also show significant suppression of the light shift for CPT-Ramsey clocks with the use of synthetic frequency protocol. The same approach can be also applied for so-called pulsed optical pumping (POP) clocks \cite{levi2012}. All these examples demonstrate an universality and efficiency of synthetic frequency protocol, which can be used in any type of clocks based on Ramsey spectroscopy.

In addition, the above analysis, taking into account the decoherence \cite{comm1}, leads us to the question about existence/absence of some hyper-Ramsey protocol, which shows zero shift $\bar{\delta}^{}_{T}$=0 for arbitrary $\Gamma$, $\Delta$, $\Omega_0$, $\tau$, and $T$. In the case of the existence, such protocol can be called as `ultimate hyper-Ramsey'.

To conclude, the synthetic frequency protocol in the Ramsey spectroscopy is a novel technique that offers a spectroscopic signal that is virtually free from probe-induced frequency shifts and their fluctuations. Our method has broad applications for any types of clocks, especially those based on ultra-narrow transitions, two-photon transitions, lattice clocks based on bosonic isotopes with controlled collision shifts \cite{aka08,lis09}, CPT-Ramsey and POP-Ramsey clocks. Moreover, our approach opens a prospect for the high-precision optical clocks based on direct frequency comb spectroscopy. High resolution matter-wave sensors \cite{yve03} are also expected to benefit from the suppression of phase shifts in the interference patterns due to the excitation pulses.

The work was supported by the Russian Scientific Foundation (project No.
16-12-00052).
M. Yu. Basalaev was supported by the Russian Foundation for Basic
 Research (Projects No. 16-32-60050 mol\_a\_dk and No. 16-32-00127 mol\_a).

\section*{Appendix}

We recalculate, with the formalism developed in \cite{zanon2015}, the approximate synthetic frequency shifts given by Eqs.~(\ref{Syn_HRA_gen}) and (\ref{Syn_HRA}) based on the Hyper-Ramsey function $n^{(e)}_{\textup{HR1}}$ from Section III [see Eq.~(\ref{n_HRA})]. The generalized Hyper-Ramsey phase $\Phi(\delta_{p})$ is expressed with laser parameters from Section II as following:
\begin{equation}
\begin{split}
\Phi(\delta_{p})=\arctan\left[\frac{\frac{\frac{\delta_{\textup{p}}}{\Omega}\tan2\theta+\frac{\delta_{\textup{p}}}{\Omega}\tan\theta}{1-\frac{\delta_{\textup{p}}^{2}-\Omega_{0}^{2}}{\Omega^{2}}\tan2\theta\tan\theta}+\frac{\frac{\delta_{\textup{p}}}{\Omega}\tan\theta+\frac{\delta_{\textup{c}}}{\Omega_{\textup{c}}}\tan\theta_{\textup{c}}}{1-\frac{\delta_{\textup{p}}\delta_{\textup{c}}}{\Omega\Omega_{\textup{c}}}\tan\theta\tan\theta_{\textup{c}}}}{1-\frac{\frac{\delta_{\textup{p}}}{\Omega}\tan2\theta+\frac{\delta_{\textup{p}}}{\Omega}\tan\theta}{1-\frac{\delta_{\textup{p}}^{2}-\Omega_{0}^{2}}{\Omega^{2}}\tan2\theta\tan\theta}
\frac{\frac{\delta_{\textup{p}}}{\Omega}\tan\theta+\frac{\delta_{\textup{c}}}{\Omega_{\textup{c}}}\tan\theta_{\textup{c}}}{1-\frac{\delta_{\textup{p}}\delta_{\textup{c}}}{\Omega\Omega_{\textup{c}}}\tan\theta\tan\theta_{\textup{c}}}}\right],
\end{split}
\label{eq:Generalized-Hyper-Ramsey-phase}
\end{equation}
including a reduced composite variable as
\begin{equation}
\begin{split}
\frac{\delta_{\textup{c}}}{\Omega_{\textup{c}}}\tan\theta_{\textup{c}}\equiv2\frac{\delta_{p}}
{\Omega}\frac{\tan2\theta\tan\theta}{\tan\theta-\tan2\theta}\,,
\end{split}
\label{reduced-variable}
\end{equation}
where $\theta=\Omega\tau/2$, $\Omega^{2}=\delta_{p}^{2}+\Omega_{0}^{2}$ and $\delta_{p}=\delta-\Delta$.

The corrected high-order clock frequency shift $\overline{\delta\omega}(T_{i})$ using different free evolution times $T_{i}$ $(i=1,2,3)$ under the condition $4\tau\ll T_{i}$ takes the explicit form:
\begin{equation}
\begin{split}
\overline{\delta\omega}(T_{i})=\frac{\Phi(\delta_{\textup{p}},\delta\rightarrow0)}
{T_{i}+\frac{\partial\Phi(\delta_{\textup{p}})}{\partial\delta}|_{\delta\rightarrow0}}\,.
\label{synthetic-frequency}
\end{split}
\end{equation}
In particular, we consider the case of $T_{1}=T$, $T_{2}=T/2$ and $T_{3}=T/3$, for which we could have the three possible synthetic clock frequencies given by:
\begin{subequations}
\begin{align}
\bar{\delta}_{T}&=\overline{\delta\omega}(T)\,,\label{eq:synthetic}\\
\bar{\delta}_\text{syn}^{(1)}&=2\overline{\delta\omega}(T)-\overline{\delta\omega}(T/2)\,,\label{eq:synthetic-1}\\
\bar{\delta}_\text{syn}^{(2)}&=3\overline{\delta\omega}(T)-3\overline{\delta\omega}(T/2)+
\overline{\delta\omega}(T/3)\,.\label{eq:synthetic-2}
\end{align}
\end{subequations}
These results given by Eqs.~(\ref{eq:synthetic})-(\ref{eq:synthetic-2}) are in very good agreement with approximated synthetic frequency shifts presented in Eq.~(\ref{Syn_HRA}) and reported in figure Fig.~\ref{HRA_1} of Section III when the condition $|\Delta/\Omega_{0}|^{2}\ll 1$ is achieved.


\begin{thebibliography}{22}

\bibitem{katori2015}
I. Ushijima, M. Takamoto, M. Das, T.~Ohkubo, and H.~Katori, Nature Photonics {\bf 9}, 185 (2015).

\bibitem{ye2015}
T. L. Nicholson, S. L. Campbell, R. B. Hutson, G.~E.~Marti, B.~J.~Bloom, R. L. McNally, W. Zhang, M. D. Barrett, M.~S.~Safronova, G.~F.~Strouse, and W.~L.~Tew, and J. Ye, Nature Communications {\bf 6}, 6896 (2015).

\bibitem{oates2013}
N. Hinkley, J. A. Sherman, N. B. Phillips, M. Schioppo, N. D. Lemke, K. Beloy, M. Pizzocaro, C. W. Oates, and A. D. Ludlow, Science {\bf 341}, 1215 (2013).

\bibitem{bize2013}
R. Le Targat, L. Lorini, Y. Le Coq, M. Zawada, J.~Guena, M.~Abgrall, M.~Gurov, P. Rosenbusch, D. G.~Rovera, B.~Nagorny, R.~Gartman, P.~G.~Westergaard, M.~E.~Tobar, M. Lours, G. Santarelli, A. Clairon, S. Bize, P. Laurent, P. Lemonde, J. Lodewyck, Nature Communications {\bf 4}, 2109 (2013).

\bibitem{ye2014}
B. J. Bloom, T. L. Nicholson, J.~R.~Williams, S.~L.~Campbell, M.~Bishof, X.~Zhang, W. Zhang, S. L. Bromley, and J. Ye, Nature {\bf 506}, 71 (2014).

\bibitem{oates2014}
K. Beloy, N. Hinkley, N. B. Phillips, J.~A.~Sherman, M.~Schioppo, J.~Lehman, A. Feldman, L. M. Hanssen, C. W. Oates, and A. D. Ludlow, Phys. Rev. Lett. {\bf 113}, 260801 (2014).

\bibitem{sterr2014}
S. Falke, N. Lemke, C. Grebing, B.~Lipphardt, S.~Weyers, V.~Gerginov, N.~Huntemann, C. Hagemann, A. Al-Masoudi, S. Hafner, S. Vogt, U.Sterr, and C. Lisdat, New Journal of Physics {\bf 16}, 073023 (2014).

\bibitem{ertmer2015}
A. P. Kulosa, D. Fim, K. H. Zipfel, S.~R$\ddot{\text{u}}$hmann, S.~Sauer, N.~Jha, K.~Gibble, W. Ertmer, E. M. Rasel, M.~S.~Safronova, U.~I.~Safronova, and S. G. Porsev, Phys. Rev. Lett. {\bf 115}, 240801 (2015).

\bibitem{huntemann2016}
N. Huntemann, C. Sanner, B. Lipphardt, Chr. Tamm, and E. Peik, Phys. Rev. Lett. {\bf 116}, 063001  (2016).

\bibitem{rosenband2010}
C. W. Chou, D. B. Hume, J.~C.~J.~Koelemeij, D.~J.~Wineland, and T. Rosenband, Phys. Rev. Lett. {\bf 104}, 070802 (2010).

\bibitem{madej2012}
A. A. Madej, P. Dube, Z. Zhou, J.~E.~Bernard, and M.~Gertsvolf, Phys. Rev. Lett. {\bf 109}, 203002 (2012).

\bibitem{gill2014}
R. M. Godun, P. B. R. Nisbet-Jones, J.~M.~Jones, S.~A.~King, L.~A.~M.~Johnson, H. S. Margolis, K. Szymaniec, S. N. Lea, K. Bongs, and P. Gill, Phys. Rev. Lett. {\bf 113}, 210801 (2014).




\bibitem{peik2003}
E. Peik and Chr. Tamm, Europhys. Lett. {\bf 61}, 181 (2003).

\bibitem{campbell2012}
C. J. Campbell, A. G. Radnaev, A.~Kuzmich, V.~A.~Dzuba,
V.~V.~Flambaum, and A. Derevianko, Phys. Rev. Lett. {\bf 108}, 120802 (2012).

\bibitem{yamaguchi2015}
A. Yamaguchi, M. Kolbe, H. Kaser, T.~Reichel, A.~Gottwald, and E. Peik, New Journal of Physics {\bf 17}, 053053 (2015).

\bibitem{tkalya2015}
E. V. Tkalya, C. Schneider, J. Jeet, and E.~R.~Hudson, Phys. Rev. C {\bf 92}, 054324 (2015).





\bibitem{derevianko2012}
A. Derevianko, V. A. Dzuba, and V.~V.~Flambaum, Phys. Rev. Lett. {\bf 109}, 180801 (2012).

\bibitem{safronova2014prl}
M. S. Safronova, V. A. Dzuba, V. V. Flambaum, U.~I.~Safronova, S.~G.~Porsev, and M. G. Kozlov, Phys. Rev.
Lett. {\bf 113}, 030801 (2014).

\bibitem{yudin2014}
V. I. Yudin, A. V. Taichenachev, and A. Derevianko, Phys. Rev. Lett. {\bf 113}, 233003 (2014).


\bibitem{ludlow2015rmp}
A. D. Ludlow, M. M. Boyd, J. Ye, E.~Peik, and P.~O.~Schmidt, Rev. of Mod. Phys. {\bf 87}, 637 (2015).





\bibitem{yudin06}
A. V. Taichenachev, V. I. Yudin, C.~W. Oates, C.~W.~Hoyt, Z.~W.~Barber, and L.~Hollberg, Phys. Rev. Lett. {\bf 96}, 083001 (2006).

\bibitem{bar06}
Z. W. Barber, C. W. Hoyt, C. W. Oates, L.~Hollberg, A.~V.~Taichenachev, and V.~I.~Yudin, Phys. Rev. Lett. {\bf 96}, 083002 (2006).

\bibitem{hos09}
K. Hosaka, S. A. Webster, A. Stannard, B.~R.~Walton, H.~S.~Margolis, and P. Gill, Phys. Rev. A {\bf 79}, 033403 (2009).

\bibitem{fis04}
M. Fischer, N. Kolachevsky, M. Zimmermann, R.~Holzwarth, Th.~Udem, T. W. H$\ddot{\text{a}}$nsch, M.~Abgrall, J.~Gr$\ddot{\text{u}}$nert, I.~Maksimovic, S.~Bize, H.~Marion, F.~Pereira Dos Santos, P.~Lemonde, G. Santarelli, P.~Laurent, A.~Clairon, C.~Salomon, M.~Haas, U.~D.~Jentschura, and C.~H.~Keitel, Phys. Rev. Lett. {\bf 92}, 230802 (2004).

\bibitem{badr06}
T. Badr, M. D. Plimmer, P. Juncar, M.~E.~Himbert, Y.~Louyer, and D. J. E. Knight, Phys. Rev. A {\bf 74}, 062509 (2006).

\bibitem{fortier06}
T. M. Fortier, Y. Le Coq, J. E. Stalnaker, D. Ortega, S.~A.~Diddams, C.~W.~Oates, and L. Hollberg, Phys. Rev. Lett. {\bf 97}, 163905 (2006).

\bibitem{stowe08}
M. C. Stowe, M. J. Thorpe, A. Pe'er, J.~Ye, J.~E.~Stalnaker, V.~Gerginov, and S. A. Diddams, Adv. At. Mol. Opt. Phys. {\bf 55}, 1 (2008).

\bibitem{yudin2010}
V. I. Yudin, A. V. Taichenachev, C.~W.~Oates, Z.~W.~Barber, N.~D.~Lemke, A. D. Ludlow, U. Sterr, Ch. Lisdat, and F. Riehle, Phys. Rev. A {\bf 82}, 011804(R) (2010).

\bibitem{hunt12}
N. Huntemann, B. Lipphardt, M. Okhapkin, Chr.~Tamm, E.~Peik, A.~V.~Taichenachev, and V. I. Yudin, Phys. Rev. Lett. {\bf 109}, 213002 (2012).

\bibitem{NPL2015}
R. Hobson, W. Bowden, S.~A.~King, P.~E.~G.~Baird, I.~R.~Hill, P. Gill, Phys. Rev. A {\bf 93}, 010501(R) (2016).

\bibitem{Zanon}
T. Zanon-Willette, E. de Clercq, and E. Arimondo, Phys. Rev. A {\bf 93}, 042506 (2016).

\bibitem{yudin2011}
V. I. Yudin, A.~V.~Taichenachev, M.~V.~Okhapkin, S.~N.~Bagayev, Chr. Tamm, E. Peik, N.~Huntemann, T.~E.~Mehlst$\ddot{\text{a}}$ubler, and F. Riehle, Phys. Rev. Lett. {\bf 107}, 030801 (2011).

\bibitem{rams1950}
N.~F. Ramsey, Phys. Rev. {\bf 78},  695  (1950).

\bibitem{tai09}
A. V. Taichenachev, V. I. Yudin, C.~W.~Oates, Z.~W.~Barber, N.~D.~Lemke, A. D. Ludlow, U. Sterr, Ch. Lisdat, and F. Riehle, JETP Lett. {\bf 90}, 808 (2009).

\bibitem{mor89}
A. Morinaga, F. Riehle, J. Ishikawa, and J. Helmcke, Appl. Phys. B {\bf 48}, 165 (1989).


\bibitem{zanon2015}
T. Zanon-Willette, V. I. Yudin, A. V. Taichenachev, Phys. Rev. A {\bf 92}, 023416 (2015).


\bibitem{zanon2005}
T. Zanon, S. Guerandel, E. de Clercq, D.~Holleville, N.~Dimarcq, and A.~Clairon, Phys. Rev. Lett. {\bf 94}, 193002
(2005).

\bibitem{chen2010}
X. Chen, G.-Q. Yang, M.-S. Wang, and J.~Zhan, Chin. Phys.
Lett. {\bf 27}, 113201 (2010).

\bibitem{blanshan2015}
E. Blanshan, S. M. Rochester, E.~A.~Donley, and J.~Kitching, Phys. Rev. A {\bf 91}, 041401(R) (2015)


\bibitem{levi2012}
S. Micalizio, C. E. Calosso, A. Godone, and F. Levi,
Metrologia {\bf 49}, 425 (2012).

\bibitem{comm1}
Note that in this paper we did not consider a spontaneus relaxation of the upper level $|e\rangle$, that can have an importance for some clock transitions. Therefore, this problem requres an addition investigation.

\bibitem{aka08}
T. Akatsuka, M. Takamoto, and H. Katori, Nature Physics {\bf 4},
954  (2008).

\bibitem{lis09}
Ch. Lisdat, J. S. R. Vellore Winfred, T.~Middelmann, F.~Riehle, and U.~Sterr, Phys. Rev. Lett. {\bf 103}, 090801 (2009).



\bibitem{yve03}
F. Yver-Leduc, P. Cheinet, J. Fils, A.~Clairon, N.~Dimarcq, D.~Holleville, P. Bouyer, and A. Landragin, J. Opt. B: Quantum Semiclass. Opt. {\bf 5}, S136 (2003).


\end{thebibliography}
\end{document}